\documentclass[twocolumn,prd,aps,groupedaddress,nopacs,nofootinbib,floats,floatfix]{revtex4}

\usepackage{color}

\usepackage{amsmath}
\usepackage{mathrsfs}
\usepackage{amssymb}

\numberwithin{equation}{section}

\usepackage{epsfig}

\newcommand{\om}{\Omega_m}
\newcommand{\ode}{\Omega_{de}}
\newcommand{\zmax}{z_{\rm max}}

\def \beq  {\begin{equation}}
\def \eeq  {\end{equation}}
\def \ber  {\begin{eqnarray}}
\def \eer  {\end{eqnarray}}

\def\mn{{Mon.\@ Not.\@ Roy.\@ Ast.\@ Soc.\ }}

\begin{document}
\newcommand{\newc}{\newcommand}

\newc{\be}{\begin{equation}}
\newc{\ee}{\end{equation}}
\newc{\ba}{\begin{eqnarray}}
\newc{\ea}{\end{eqnarray}}
\newc{\bea}{\begin{eqnarray*}}
\newc{\eea}{\end{eqnarray*}}
\newc{\D}{\partial}
\newc{\ie}{{\it i.e.} }
\newc{\eg}{{\it e.g.} }
\newc{\etc}{{\it etc.} }
\newc{\etal}{{\it et al.}}
\newc{\lcdm }{$\Lambda$CDM }
\newcommand{\nn}{\nonumber}
\newc{\ra}{\rightarrow}
\newc{\lra}{\leftrightarrow}
\newc{\lsim}{\buildrel{<}\over{\sim}}
\newc{\gsim}{\buildrel{>}\over{\sim}}


\title{Cosmographic Degeneracy}
\author{Arman Shafieloo$^1$ and Eric V.\ Linder$^{1,2}$}
\affiliation{$^1$ Institute for the Early Universe WCU, Ewha Womans 
University, Seoul, 120-750, Korea} 
\affiliation{$^2$ Berkeley Lab \& University of California, Berkeley, CA 
94720, USA}

\date{\today}

\begin{abstract} 
We examine the dark energy and matter densities allowed by precision 
measurements of distances out to various redshifts, in the presence of spatial 
curvature and (near) arbitrary behavior of the dark energy equation of state. 
Degeneracies among the parameters permit a remarkably large variation in 
their values when using only distance measurements of the late time 
universe and making no assumptions about the dark energy or curvature.  Going 
beyond distance measurements to a lower limit on the growth of structure 
bounds the allowed region significantly but still leaves considerable 
freedom to match a flat $\Lambda$CDM model with dark energy very different 
from a cosmological constant.   The combination of distances 
with Hubble parameter, gravitational lensing or other large scale structure 
data is essential to determining robustly the cosmological model. 
\end{abstract}
\maketitle

\section{Introduction}

A central goal of modern cosmology is to reveal in detail the energy 
budget of the universe.  In addition to matter (baryonic and dark matter) 
and small contributions by radiation and neutrinos, there is an 
(effective) dark energy associated with the accelerated expansion and 
possibly an (effective) curvature energy associated with deviation from 
spatial flatness.  Cosmological observations, especially over the past 
decade, have made great strides in constraining the energy density fractions 
in each of these, but generally assuming specific behaviors for the dark 
energy.  Since dark energy is an almost total mystery, it behooves us to 
reexamine the issue and ask to what extent the $\Lambda$CDM concordance model 
of $\sim28\%$ matter density and $\sim72\%$ cosmological constant energy 
density is necessarily close to the true energy budget. 

Because all the energy densities enter into the Hubble expansion rate, 
which then determines the distance-redshift relation, degeneracies exist 
between the components such that more of one can compensate for less of 
another.  Since they evolve differently with redshift, however, each 
characterized by their own equation of state parameter (0 for matter, 
$-1/3$ for curvature, $w(z)$ for dark energy), one expects that observations 
over a sufficiently wide redshift range give leverage to break the 
degeneracies.  This has been explored for restricted scenarios 
of matter and dark energy densities (e.g.\ \cite{steinhardt,Shafieloo05,Shafieloo07,om}) and curvature 
and dark energy densities (e.g.\ \cite{lincurv,bassett}), and 
nonparametrically from the observations through redshifts bins of dark 
energy (e.g.\ \cite{union21}).  Perhaps closest in philosophy to our 
approach are the works of \cite{mhh1,mhh2,mhh3}, which look at how 
the diversity of models translate to dispersion in observables, while 
we explore the converse of how a tight observable relation can arise 
from a wide range of models. 

In this paper we investigate the freedom around the concordance model 
caused by degeneracies when we allow for matter, curvature, and dark 
energy with no a priori restriction on its equation of state.  The 
philosophy is to use as direct measurements as possible without making 
assumptions about the dark energy.  Therefore we restrict ourselves to 
late time observations since we have no knowledge of dark energy 
behavior at early times, e.g.\ is there early dark energy affecting the 
cosmic microwave background (CMB).  We use purely geometric distance 
measurements, examining how the degeneracies are broken as the data 
quality and redshift range improves.  That is, suppose we have even 
exact agreement with a particular flat $\Lambda$CDM model in the 
distance-redshift relation out to some redshift $z$; how close to the 
concordance model in the matter density--dark energy density plane does 
this restrict the cosmological model when allowing for curvature and 
arbitrary $w(z)$ behavior? We should note that in this paper we are considering the degeneracy between the fundamental cosmological quantities namely matter density, curvature and the effective equation of state of dark energy. There are also another sort of degeneracies between different cosmological models which are very different by nature but they result to similar effective equation of state of dark energy~\cite{SahniShtanov1,SahniShtanov2}. 

Section \ref{sec:distance} lays out the methodology and explains the 
constraints imposed by various levels of conditions imposed by consistency 
and by the observations.  The influence of measurements apart from 
distance, such as age of the universe and the linear growth factor of 
cosmic structure, are also addressed.  In Section~\ref{sec:degen} we 
quantify the constraints in the matter--dark energy density ($\om$--$\ode$) 
plane and Section~\ref{sec:concl} 
summarizes the results about how well we actually know our cosmological 
model.

\section{Cosmographic Constraints \label{sec:distance}} 

The method of our analysis is simple and straightforward.  We calculate the 
luminosity distance--redshift relation for a flat $\Lambda$CDM model with 
$\om=0.28$ and assume that (future) observations determine that distances 
agree with this model to a certain precision for all redshifts out to some 
$\zmax$.  Initially we take the agreement to be exact, to illustrate the 
level of degeneracies that persist even in this ideal case.  For our 
comparison to theoretical cosmological models we stay within the 
Friedmann-Robertson-Walker framework, taking noninteracting components 
of nonrelativistic matter, spatial curvature, and dark energy.  Since the 
dark energy is allowed to behave (nearly) arbitrarily, many elaborations 
such as interacting components can actually be folded into the dark energy 
behavior. 

The Hubble expansion parameter, giving the logarithmic time variation of 
the scale factor $a=1/(1+z)$, is 
\beq 
\begin{split} 
h(z)^2&\equiv[H(z)/H_0]^2\\ 
&=\om (1+z)^3 + (1-\om-\ode)(1+z)^2\\ 
&\quad +\ode\,\exp\left[3\int^z_0 \frac{dz'}{1+z'}\, [1+w(z')]\right] \,,\label{eq:hz}
\end{split} 
\eeq 
where the last term on the second line represents the spatial curvature 
energy density.  This leads to the luminosity distance through 
\beq 
d_l(z)=\frac{1+z}{\sqrt{1-\om-\ode}}\sinh\left(\sqrt{1-\om-\ode}\int_0^z \frac{dz'}{h(z')}\right)\,. \label{eq:dl}
\eeq 
Note that sinh is an analytic function valid for positive, zero, or negative 
curvature.  Although we write the luminosity distance, all the results 
still hold if one uses angular diameter distance $d_a=d_l/(1+z)^2$ instead. 

We concentrate on what values of the matter density $\om$ and dark energy 
density $\ode$ are allowed, without assuming a specific equation of state 
$w(z)$, given the distances $d_l(z)$.  Note that only late time parameters 
enter, i.e.\ we do not have to make any assumptions about conditions at 
high redshift in the early universe. 
 
\subsection{Radius of Curvature Condition} 

The first constraint on the $\om$--$\ode$ plane is the requirement that 
for a positive curvature universe the radius of curvature is large enough 
to allow the luminosity distance to match that of the input $\Lambda$CDM 
model out to $\zmax$.  Mathematically this corresponds to the sine 
(analytic continuation of sinh in Eq.~\ref{eq:dl}) function having an 
amplitude bounded by 1.  Basically $\om+\ode-1\le [(1+z)/d_l(z)]^2$ for 
all $z\le\zmax$.  We call the region of $\om$--$\ode$ space 
violating this ``Forbidden Region 1''.  

Note that this is a more restrictive 
condition than the ``no bounce'' condition, requiring that a transition from 
contraction to expansion is avoided out to the $\zmax$ considered.  A 
bounce -- Hubble parameter going to zero -- imposes a minimum scale factor 
and hence maximum redshift; note that when $H\to0$ then the argument of 
the sine function goes to infinity.  
Avoiding a bounce out to infinite redshift is necessary for 
having a Big Bang, but again our radius of curvature condition does not 
require any extrapolation to early universe conditions.

\subsection{Positive Density Condition} 

The next constraint on the densities comes from requiring that the dark 
energy density be positive.  One can rewrite Eq.~(\ref{eq:hz}) as 
\beq 
\ode(z)=h(z)^2-\om (1+z)^3-(1-\om-\ode)(1+z)^2 \,, 
\eeq 
and so the condition that $\ode(z)\ge 0$ for all $z\le\zmax$ restricts the 
allowed region of the density plane.  We call this ``Forbidden Region 2''.  
Note that we do not restrict $w$ from going to $-\infty$, needed to allow 
the dark energy density to go to zero.  
A negative dark energy density could also 
violate the radius of curvature condition so we expect overlap between 
the forbidden regions. 

\subsection{Age Condition} 

The above two conditions are basically consistency restrictions within the 
framework of the distance matching.  We could ask for further observational 
conditions that are general enough not to require assumptions about the 
nature of the components.  

The lookback time--redshift relation is one possibility.  It is solely a 
function of the Hubble parameter, 
\be
T(z)=t_0-t(z)= H_0^{-1} \int_{0}^{z}\frac{dz'}{(1+z')\,h(z')} \,. 
\ee 
However, we currently do not have any robust, model independent, and 
accurate limits on the lookback time--redshift relation.  One could use 
the age of the universe, the limit of lookback time as redshift gets large. 
This might seem to contradict our approach of not requiring any knowledge 
of early universe conditions but this can be worked around.  The 
observational constraint itself can be taken from the age of globular 
clusters \cite{krauss} or white dwarf stars \cite{age1}.  We consider 
these as placing a lower bound on the age of the universe of 
$12.5$ Gyr.  As far as the theoretical models, 
we simply take the maximum possible age, i.e.\ we maximize the contribution 
for redshifts above our $\zmax$ for each $\om$--$\ode$ point and if the 
total age still falls below the observational bound then that region of 
the density plane is ruled out. 

To maximize the age for a given $\om$--$\ode$ we minimize $H(z)$ at the 
high redshifts.  The minimum occurs by driving the dark energy density 
contribution to zero immediately after $\zmax$(we do not allow negative 
energy density), thus leaving only the 
matter density and curvature contributions set by the low redshift 
behavior.  In this way we obtain a robust bound that does not actually 
require knowledge of the high redshift universe conditions.  We call 
regions of the density plane that even so do not achieve the minimum 
age of 12.5 Gyr ``Forbidden Region 3''.  

Note that because this is a conservative bound, many dark energy models 
in the still-allowed region of the density plane will have a low age, but 
we only rule out the region where {\it no\/} possible dark energy behavior 
permits a universe consistent with the age constraint.  Also note that to 
estimate age we need to know the value of $H_0$; the region shown uses 
the $1\sigma$ lower limit (so as to maximize the age) of the measurement 
$H_0=73.8 \pm 2.4$ km/s/Mpc \cite{hubble}.  If we use the $2\sigma$ lower 
limit of 69, however, then no region beyond that of Forbidden Region 2 is 
ruled out for the $\zmax$ we consider.

\subsection{Growth Condition} 

While the positive dark energy density condition is very effective at 
ruling out high matter densities (since this gives a large Hubble parameter 
and hence too small distances relative to $\Lambda$CDM, and further dark 
energy density only exacerbates the situation), very low matter densities 
are allowed.  We could impose a condition that the matter density must be 
at least as large as the baryon density implied by primordial nucleosynthesis 
or the CMB, i.e.\ $\om\gtrsim0.04$, but this would use early universe 
conditions.  Instead we look at growth of structure.  Growth is particularly 
important because we allow the dark energy to behave as matter with 
equation of state $w=0$, so it enters equivalently to matter in the Hubble 
parameter (thus allowing low $\om$), but growth is also sensitive to the 
clustered density while we 
take dark energy to be smooth.  Growth thus allows distinction between them 
and breaking of the degeneracy. 

Components other than clustered matter have two effects on growth of matter 
structures: they change the expansion rate, and hence the friction term in 
the linear density perturbation growth equation, and they change the source 
term (basically the matter density).  We can write the growth equation as
\be
\delta" + (2-q)a^{-1}\delta' - (3/2)a^{-2}\,\Omega_m(a) \delta = 0 \,, 
\label{eq:grow} 
\ee 
where $q$ is the deceleration parameter and a prime denotes a derivative 
with respect to scale factor $a$.  
We will be interested in how much the growth is enhanced or suppressed 
relative to a pure matter universe, where $\delta$ grows linearly with $a$. 
A component with negative equation of state both increases the friction and 
decreases the source (through reducing $\Omega_m(a)$) relative to this case.  
One might think that a component with $w>0$ could enhance growth since it 
reduces the friction, but we will find that the reduction in the source 
term is more important. 

The presence of dark energy and curvature reduces $\Omega_m(a)$ below the 
pure matter value of $1$, and changes the friction term 
\be
2-q = (3/2)(1-w_{\rm mix}(1-\Omega_m(a)) \,. 
\ee 
where $w_{\rm mix}$ is the effective equation of state of the combination 
of curvature and dark energy.  The question is whether the reduced friction 
can make up for the reduced source term.  


Putting in the function $\delta\propto a^m$ to the growth equation 
(\ref{eq:grow}), we get the characteristic equation 
\be 
2m^2 + [1-3w_{\rm mix}(1-\Omega_m(a))]m - 3\Omega_m(a) = 0 \,. 
\ee 
For any allowed value of $\Omega_m(a)$ there is no root with $m>1$, 
i.e.\ growth can never be greater than in the pure matter case, when 
$w_{\rm mix}\le 1$.  Taking $m=1+\epsilon$, one finds 
$\epsilon=-3(1-w_{\rm mix})(1-\Omega_m(a))/(5-3w_{\rm mix}(1-\Omega_m(a)))$, 
which is always negative for $w_{\rm mix}<1$, since additional high redshift 
components can only reduce $\Omega_m(a)$ below the pure matter value of 1, 
our baseline.

Thus, the reduced source term always suppresses 
growth more than the easing from the less positive friction term.  If we 
allow $w_{\rm mix}> 1$, however, then the friction goes negative and 
growth can actually be enhanced.  So our growth condition will only apply, 
unlike the other conditions, to those dark energy models with $w\le 1$ 
at $z>\zmax$ (as all canonical scalar fields obey).  We emphasize that 
we still allow $w$ to take any value at $z\le\zmax$. 

Now we apply an argument similar to the age condition.  Since growth can 
only be suppressed upon allowing for dark energy (with $w\le1$ at $z>\zmax$) 
or curvature, we take the growth factor at $z>\zmax$ to be the 
maximal (Einstein-de Sitter pure matter) value $g=\delta/a=1$.  Then we 
calculate the growth factor for today, using the expansion history for 
any particular point in the density plane derived to match the distance 
relation to $\zmax$.  This procedure maximizes the total growth to the 
present (in fact it is overly conservative, but has the virtue of model 
independence).  
If the growth factor today is still too small to be 
acceptable observationally, which we take as $g_0 < 0.65$, or basically 
$\sigma_8 < 0.7$ (cf.~\cite{seljak}), then we consider that region as 
ruled out.  We call such an area ``Forbidden Region 4''.  As 
expected, this rules out very low matter densities.

\section{Breaking Degeneracies \label{sec:degen}} 

Having established four conditions on acceptable values for the 
density parameters despite spatial curvature and (nearly) arbitrary 
dark energy behavior, we now apply them individually and jointly to 
the density plane of $\om$--$\ode$. 

\begin{figure}[!htb]
\vspace{-0.in}
\includegraphics[width=\columnwidth, angle=0]{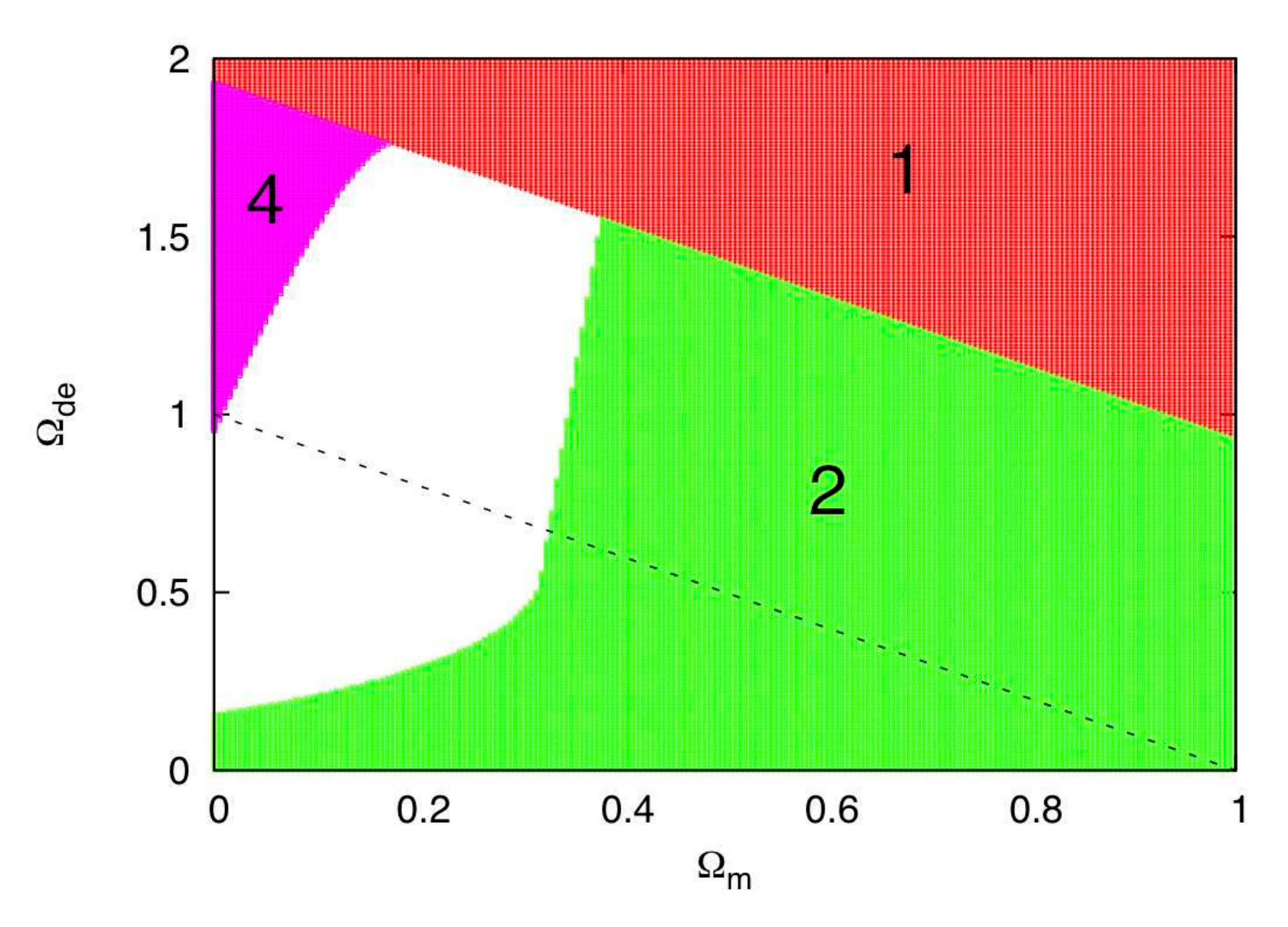}
\vspace{-0.01in}
\vspace{-0.01in}
\includegraphics[width=\columnwidth, angle=0]{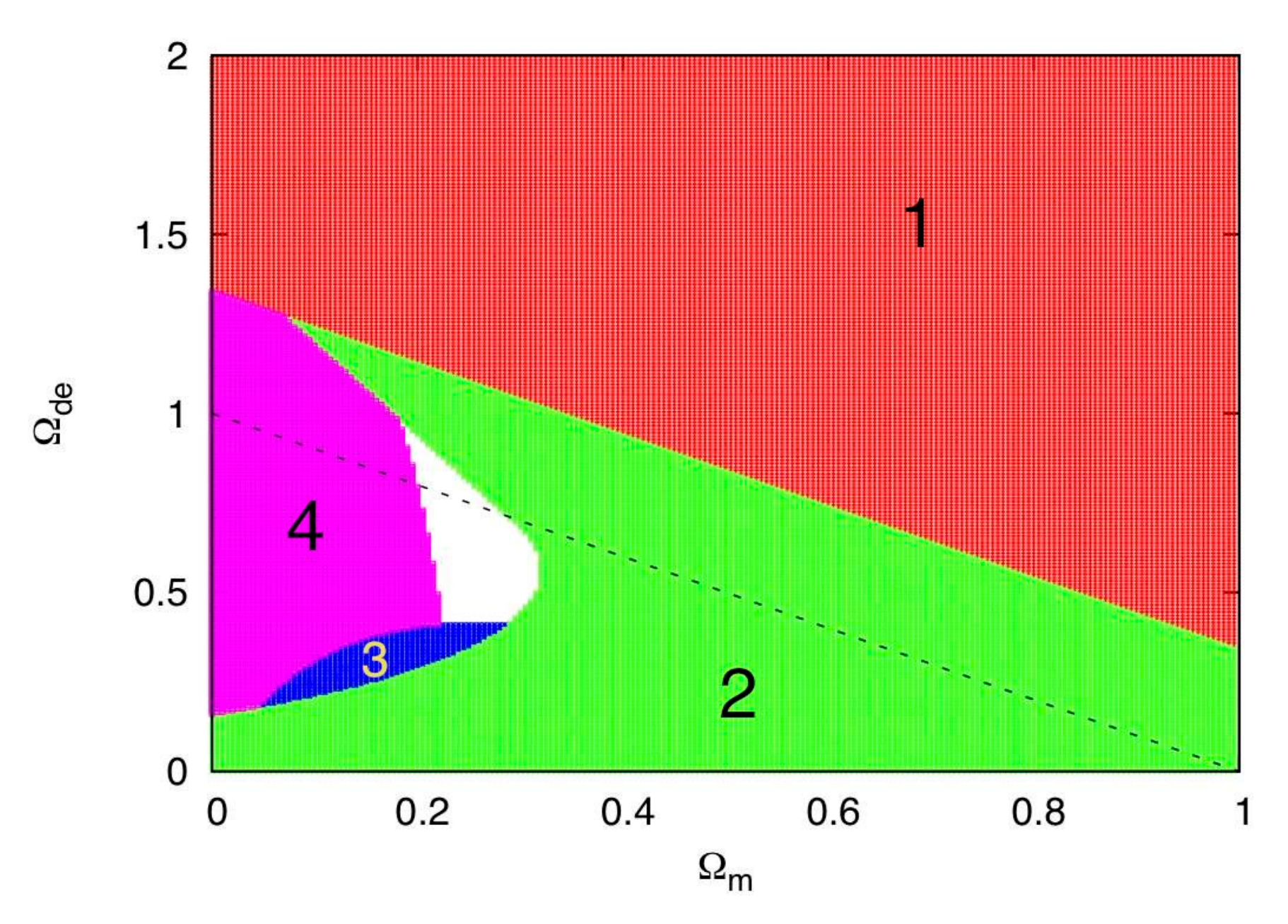}
\vspace{-0.1in}
\caption{The density parameter space $\om$--$\ode$ has forbidden 
regions from (1) minimum radius of curvature, (2) positivity of dark 
energy density, (3) minimum age of the universe, and (4) minimum total 
growth of structure.  The dotted line passes through the points where 
the spatial curvature is zero.  Parameter values lying in the unshaded 
areas are capable of exactly matching the distance-redshift relation of 
flat $\Lambda$CDM with $\om=0.28$ out to $\zmax=1.5$ (top panel) or 
$\zmax=4$ (bottom panel).} 
\label{fig:basic}
\end{figure}

Figure~\ref{fig:basic} shows the $\om$--$\ode$ parameter space and each 
of the forbidden regions.  Forbidden Region 1 from the radius of curvature 
condition constrains high positive curvature, where the sum $\om+\ode$ is 
large.  As $\zmax$ increases the maximum $d_l/(1+z)$ in the inverse of the 
sine function increases monotonically and so the sum of the energy 
densities in the curvature factor is more tightly restricted, causing 
the forbidden region boundary to sweep down through the plane.

Forbidden Region 2 from the dark energy density positivity condition 
rules out high values of $\om$, since then the Hubble parameter is so 
large (and any dark energy can only add to it) that the distances are 
too small to match the required $\Lambda$CDM values.  Similarly, in the 
negative curvature region the curvature contribution to the Hubble 
parameter is too large, especially for small values of the dark energy 
that increase the curvature contribution (as do small values of the 
matter density, but those also give a smaller matter contribution), so 
again the distances are too small to be viable.  Thus Region 2 is 
roughly (reverse) L-shaped.  As $\zmax$ increases, the positive 
curvature region becomes more restricted as the distance is further 
reduced by the curvature, i.e.\ $\sin x<x$, so the L-shape begins to 
fold over on itself. 

The minimum age of the universe, defining Forbidden Region 3, does not 
have a strong effect. Some of the higher values of $\ode$, which would 
only help to avoid Forbidden Region 2, here increase $H$ enough to 
decrease the age of the universe below the required limit.  
However, for low $\zmax$ this does not occur since the dark energy density 
can vanish at higher redshifts, and even for $\zmax=4$ the values  
$H_0<69$ km/s/Mpc cause this constraint to recede within Forbidden Region 2. 

Finally the growth condition, as predicted, gives Forbidden Region 4 
cutting off the low matter densities and high dark energy densities 
(which aid in suppressing growth).  As $\zmax$ increases, the period 
of helpful, maximal (Einstein-de Sitter) growth is shortened and so 
even low dark energy densities do not allow low matter densities to 
deliver sufficient growth. 

We reiterate the meaning of the remaining allowed region: it 
illustrates the 
cosmographic degeneracy, that for such combinations of $\om$ and 
$\ode$ there exists a combination of curvature and an evolving dark 
energy model (where the exact necessary $w(z)$ can be reconstructed) 
that results in a distance-redshift relation out to $\zmax$ exactly 
equal to that of the flat $\Lambda$CDM model with $\om=0.28$.  
So no matter how precise our measurements of $d_l(z)$, these models 
cannot be distinguished, and in particular one cannot conclude that 
the dark energy arises from a cosmological constant. Furthermore, such 
a cosmology is viable in the sense that it obeys the lower limits from 
the growth factor and the age of the universe. 

As $\zmax$ increases, the parameter region capable of matching 
exactly the distance-redshift relation of flat $\Lambda$CDM with 
$\om=0.28$ out to $\zmax$, by using the presence of curvature and 
arbitrary dark energy equation of state behavior, diminishes.  
Forbidden Region 1 sweeps down, Region 2 closes in from the right and 
above, and Region 4 squeezes from the left and below.  The top panel 
of Fig.~\ref{fig:basic} using $\zmax=1.5$ is roughly an idealized 
version of near future cosmological observations, and $\zmax=4$ might 
represent further future observations using, for example, neutral 
hydrogen surveys.  Only as $\zmax$ gets very large does the allowed 
parameter space zero in on the input flat $\Lambda$CDM, $\om=0.28$ 
model, completely breaking the cosmographic degeneracy. 

As we have already seen from use of the growth factor, other 
observations besides the distance to some redshift $z$ can prove 
effective at breaking the cosmographic degeneracy.  For example, 
distance intervals such as those entering in gravitational lensing, 
Alcock-Paczynski effect, or Hubble parameter measurements 
involve the curvature in a distinct way \cite{bernstein,knox,kazin}.  
Further measures of growth, such as relative growth between two 
redshifts or the integrated Sachs-Wolfe effect, could prove useful, 
as would direct estimate of the matter density through the clustering 
statistics of large scale structure, though these would have to allow 
for arbitrary dark energy evolution.  If we are willing to restrict to 
models where dark energy has no role at high redshifts, then the cosmic 
microwave background provides a rich source of constraints, but that is 
not the philosophy adopted here. 

\begin{figure}[!htb]
\vspace{-0.8in}
\includegraphics[scale=0.40, angle=0]{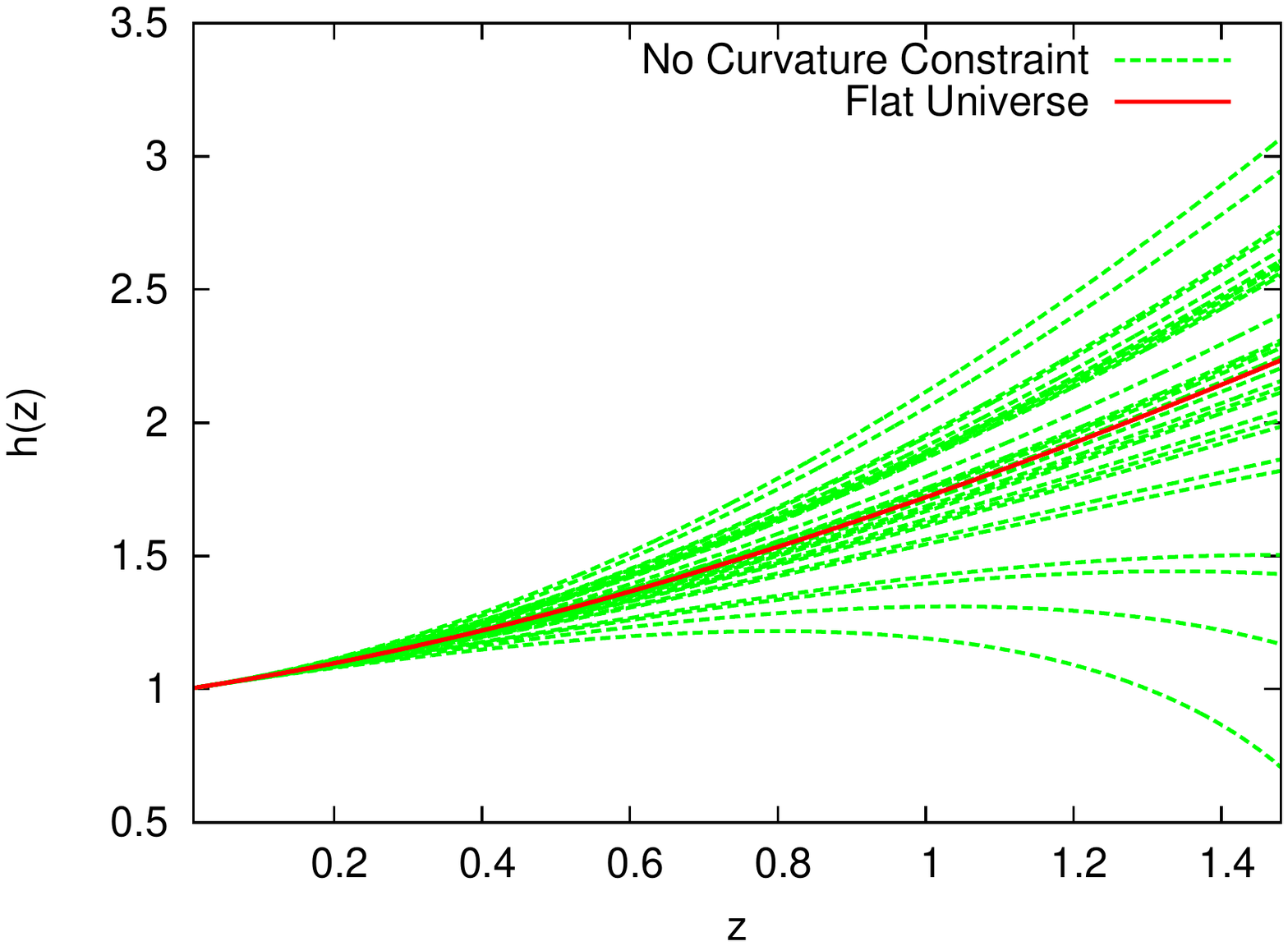}
\vspace{-0.5in}
\vspace{-0.9in}
\includegraphics[scale=0.40, angle=0]{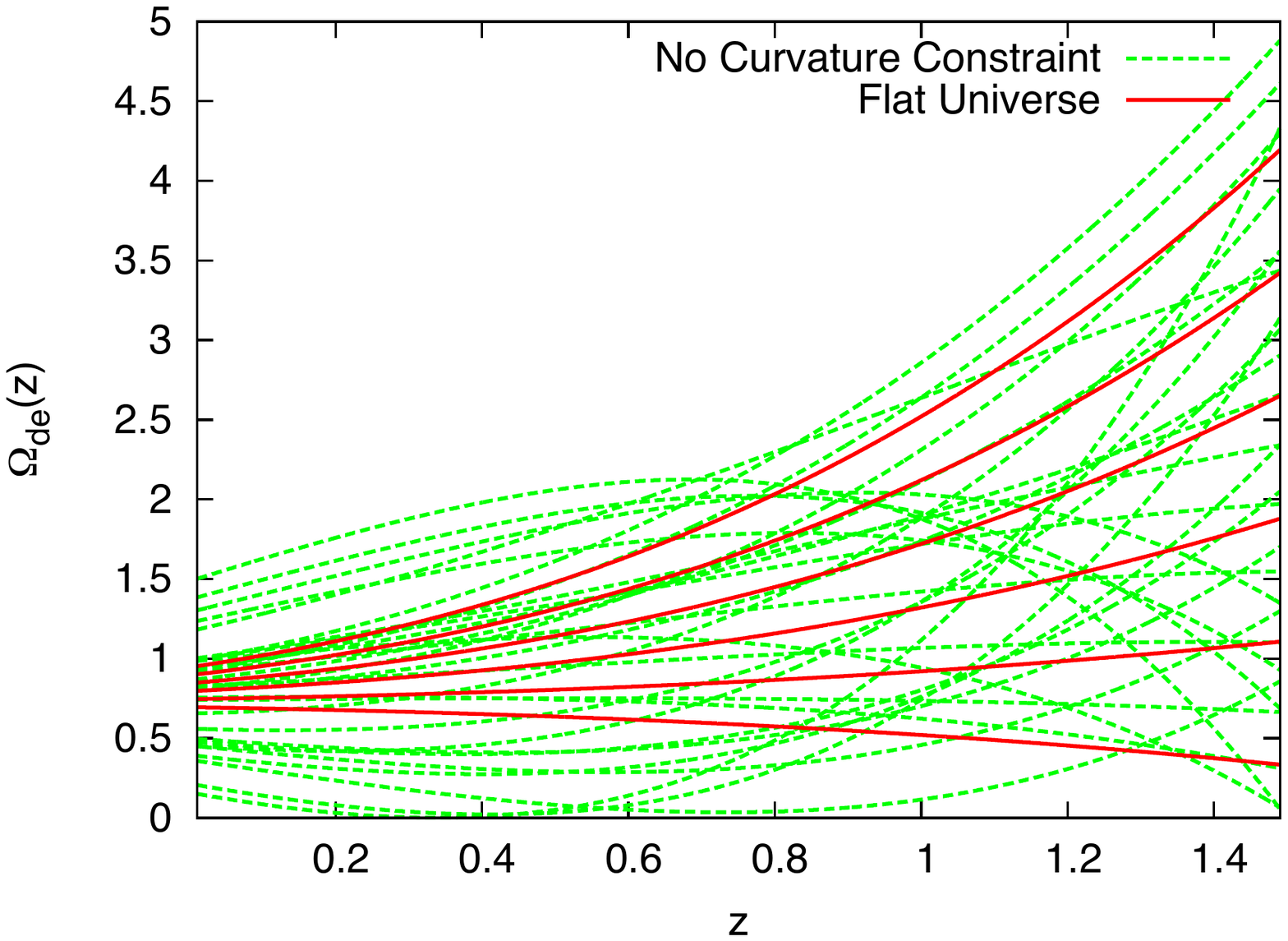}
\vspace{-0.9in}
\vspace{-0.5in}
\includegraphics[scale=0.40, angle=0]{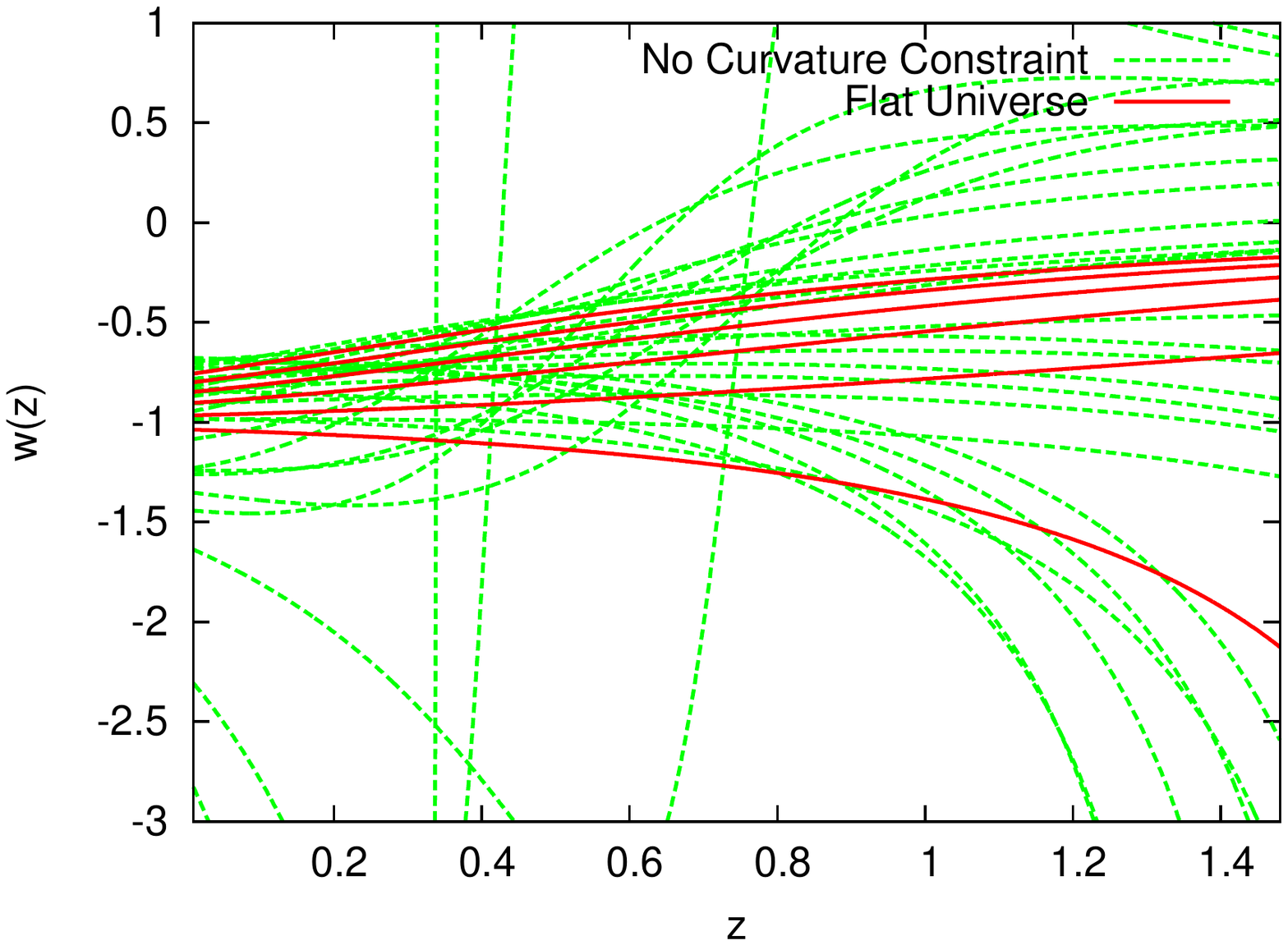}
\vspace{-0.5in}
\caption{Non-exhaustive sample of the Hubble parameter $h(z)$, dark energy 
density $\Omega_{de}(z)$, and $w(z)$ for different points in the 
$\Omega_{m}$-$\Omega_{de}$ parameter space that match the $\Lambda$CDM 
distances exactly out to $\zmax=1.5$ and satisfy our four conditions.  
Light green lines represent the results with curvature allowed to be nonflat 
and dark red lines restrict to the zero curvature case. The assumed true model 
is a spatially flat $\Lambda$CDM model with $\om=0.28$.} 
\label{fig:cosmoquan}
\end{figure}

Figure~\ref{fig:cosmoquan} illustrates non-exhaustively the variety 
of Hubble parameters $h(z)$, dark energy density $\Omega_{de}(z)$, 
and dark energy equations of state 
$w(z)$ that exist at different points in the $\om$--$\ode$ density space 
that satisfy all our mathematical and cosmological constraints, out to 
$\zmax=1.5$.  Light green lines represent results when allowing for 
curvature as well as dark energy behavior, while dark red lines fix 
curvature to be zero and only employ the freedom in the dark energy 
equation of state.

For the case of $h(z)$, models with the same 
curvature, and forced to have the same distances, necessarily have the 
same $h(z)$ (think of it as a derivative of the distance).  For 
$\Omega_{de}(z)$, however, there is additional freedom coming from 
changing the matter contribution even 
if the curvature is fixed, and this carries through to $w(z)$ as well.  
Without fixing curvature an even greater variety of dark energy 
behaviors is exhibited -- despite the distance relation agreeing 
perfectly with a cosmological constant universe. 

To this point we have required exact distance matching, i.e.\ perfect 
precision on the observations that agree with the $\Lambda$CDM model. 
Even so, the cosmographic degeneracy region is substantial.  We now 
consider the effect of some uncertainty (statistical or systematic) 
in the distance measurements by accepting as an allowed region in the 
density plane any values that fit a flat $\Lambda$CDM model with matter 
densities $0.26\le\om\le0.3$.  This roughly corresponds to a 1.6\% 
(2.3\%) distance uncertainty out to $\zmax=1.5$ (4). 

\begin{figure}[!htb]
\vspace{-.0in}
\includegraphics[width=\columnwidth,angle=0]{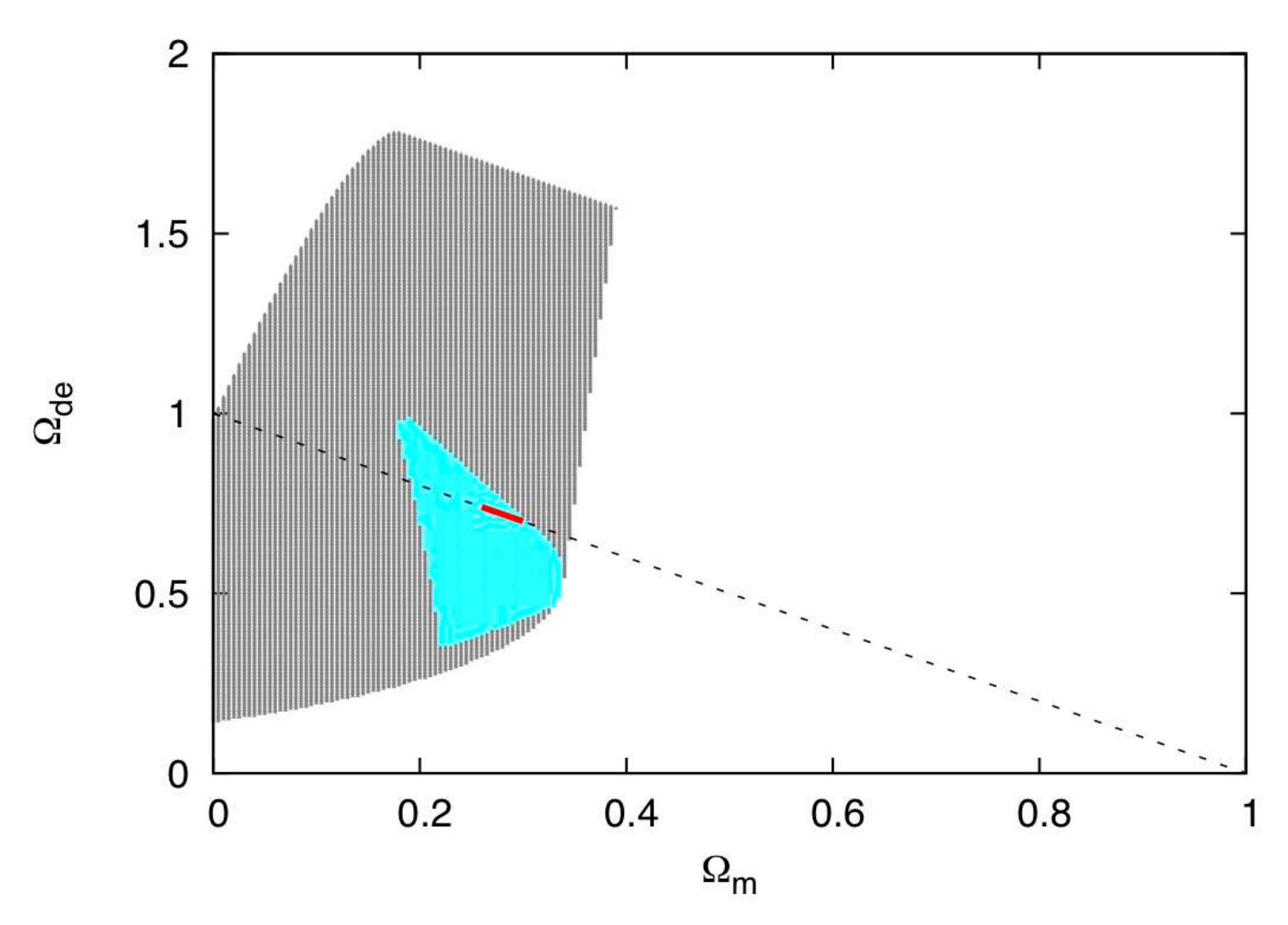}
\vspace{-0.1in}
\caption{Allowed region in the density parameter space if the true model can 
be any flat $\Lambda$CDM model with $0.26\le\om\le0.3$, and the four 
forbidden regions are excluded.  The grey area applies the distance 
matching up to $\zmax=1.5$ and the blue light area to $\zmax=4$.  The black 
dashed line shows the flatness line and the short red solid line along it 
represents the range of true, $\Lambda$CDM models allowed. 
} 
\label{fig:error}
\end{figure}

Figure~\ref{fig:error} shows the allowed region in the density parameter 
space now allowing for this uncertainty in the true distances, and applying 
the forbidden regions as before.  The larger, grey shaded area 
is for $\zmax=1.5$ and the smaller, blue light area is for $\zmax=4$.  The black 
dashed line shows the flatness line and the short solid red line shows the range of true flat $\Lambda$CDM models.  We see that 
the regions allowed with such a variation in true distances does not 
expand greatly over the white, unshaded regions allowed in 
Fig.~\ref{fig:basic}, so this level of measurement uncertainty does not 
change our conclusions.

\section{Conclusions \label{sec:concl}}

We have quantified the cosmographic degeneracy present when matter, 
spatial curvature, and unrestricted dark energy models can contribute 
to the distance-redshift relation.  Even when perfectly matching the 
distances out to $\zmax=1.5$ for a flat $\Lambda$CDM model with a given 
matter density, a substantial region of the density parameter space remains 
degenerate with the true model.  This implies that we {\it cannot\/} 
assume that the cosmological constant describes the dark energy through 
such distance measurements alone.  

Imposing other low redshift constraints, such as basic consistency 
conditions on the radius of curvature of closed universes and positivity 
of the dark energy density, and observational criteria such as a minimum 
age of the universe and a simple lower bound on the total growth factor 
for large scale structure, still leaves considerable freedom for the 
curvature and dark energy contributions.  One would zero in on a 
$\Lambda$CDM model only when restricting the form of the dark energy 
evolution or bringing in early universe constraints (e.g.\ CMB and more 
detailed large scale structure characteristics) -- assuming dark energy 
has negligible contribution at high redshift.  Distances involving 
curvature differently, e.g.\ parallax \cite{weinberg} or distance 
intervals, including direct measurement of the Hubble parameter or 
gravitational lensing, of galaxies or the CMB, may offer another path 
(see e.g.\ \cite{bernstein,knox,kazin,suyu}). 

We emphasize that we have made essentially no assumptions about the 
dark energy behavior, allowing its equation of state parameter $w$ to 
range from $-\infty$ to $+\infty$ (with the exception of restricting 
$w\le1$ above $\zmax$ when using the minimum growth condition).  A 
constraint on $w$ to lie within $[-1,+1]$, say, the values for a 
canonical scalar field, would 
limit the allowed region to a much smaller area around the true values in 
the density space. When the assumed $\Omega_{m}$ is larger than the actual 
matter density, at some redshifts we need a very large negative equation 
of state of dark energy to suppress the contribution of dark energy in the 
total energy density of the universe.  Conversely, when the assumed 
$\Omega_m$ is smaller than the actual matter density we need a dark 
energy with equation of state of greater than $-1$ and in some cases 
even greater than $1$ to compensate the lack of contribution from the 
matter part.  Thus, imposing the limit $w < +1$, say, rules out an area 
with low matter and dark energy densities.  
A constraint on spatial curvature (while still allowing for arbitrary 
dark energy behavior) would cut a diagonal swath in the density plane 
parallel to the flatness line, considerably restricting the allowed 
region. 

It is interesting to see how our ignorance of the nature of dark energy 
and the geometric curvature of space diffuse the strength of evidence 
for the cosmological constant model from distance measurements.  The 
true universe may be much more complicated, and yet perfectly consistent 
with cosmography, than this highly restricted model.  Combining 
distance measurements with gravitational lensing data and mapping of 
large scale structure will greatly reduce the 
degeneracies exhibited and such a suite of future observations offers 
true hope to understand our universe in a less model dependent manner.

\acknowledgments

This work has been supported by World Class University grant 
R32-2009-000-10130-0 through the National Research Foundation, Ministry 
of Education, Science and Technology of Korea and in part by the Director, 
Office of Science, Office of High Energy Physics, of the U.S.\ Department 
of Energy under Contract No.\ DE-AC02-05CH11231.

\end{document}